# Nitrate Deposition following an Astrophysical Ionizing Radiation Event

Ben Neuenswander, Specialized Chemistry Center, University of Kansas, Lawrence, KS 66047, USA

Adrian Melott, Department of Physics and Astronomy, University of Kansas

Abstract

It is known that a gamma ray burst (GRB) originating near the Earth could be devastating to life. The mechanism of ozone depletion and subsequent increased UVB exposure is the primary risk, but models also show increased nitrification culminating in nitric acid rainout. These effects are also expected after nearby supernovae and extreme solar proton events. In this work we considered specifically whether the increased nitric acid rainout from such events is a threat to modern terrestrial ecosystems. We also considered its potential benefit to early terrestrial Paleozoic ecosystems. We used established critical loads for nitrogen deposition in ecoregions of Europe and the US and compared them with previously predicted values of nitric acid rainout from a typical GRB within our galaxy. The predicted rainout was found to be too low to harm modern ecosystems, however, it is large compared with probable nitrate flux onto land prior to the invasion of plants. We suggest that this flux may have contributed nutrients to this invasion if, as hypothesized, the end-Ordovician extinction event were initiated by a GRB or other ionizing radiation event.

Keywords

gamma ray burst; supernova; solar proton event; nitrate deposition; late-Ordovician

[1]bennder@ku.edu, (785)312-1108

1. Introduction

Bursts of ionizing radiation episodically alter the Earth's atmosphere.  Sources include gamma ray bursts (GRBs), supernovae, and probably the Sun (Melott and Thomas, 2011, Thomas et al., 2013), although recent claims of numerous recent supernovae are probably incorrect (Melott et al., 2014).  Astrophysical ionizing radiation can provide sufficient energy to break the normally stable $N_2$ triple bond which facilitates the formation of so-called "odd nitrogen" compounds ($NO_x$). $NO_x$ can, among other things, react with ozone, catalyzing its conversion back to molecular oxygen. As stratospheric ozone is consumed, additional solar ultraviolet-B (UVB) radiation passes to the surface resulting in increased mortality of exposed organisms, particularly phytoplankton despite their ability to avoid and repair UV damage (Melott and Thomas, 2011). This process has been hypothesized as a link to mass extinctions, particularly the end-Ordovician extinction, which fits some of the patterns expected for such an event (Melott et al., 2004, Melott and Thomas, 2009). A GRB of sufficient luminosity is expected, based on rates, to have occurred multiple times during the history of life on Earth.  However, due to very different atmospheric chemistry and the possible lack of a UVB shield  (Melott et al., 2005, Piran and Jimenez, 2014), the ozone depletion mechanism is probably largely irrelevant prior to the advent of free oxygen.

The presence of $NO_x$ in the atmosphere also eventually leads to increased nitrate rainout in the form of nitric acid ($HNO_3$) (Thomas et al., 2005). This has both the potential to harm or benefit life.  Life requires that nitrogen be oxidized or reduced from its gaseous form ($N_2$), called nitrogen fixation, before it is available for biologic use. Prior to the Haber-Bosch ammonia manufacturing process which began in the early 1900's most nitrogen fixation was done by biologic sources in the form of nitrogenase-containing bacteria and archaea, with much lower contributions from abiotic sources which include lightning, volcanism and bolide impacts (Schlesinger, 1997, Galloway et al., 2004).  Astrophysical ionizing radiation events produce intermittent doses of nitrate flux in addition to normal biotic and abiotic sources, as shown specifically for a GRB by Thomas et al., (2005). Although nitrogen fixation is necessary for life, it is known that excess nutrient deposition, or eutrophication, can be harmful (Galloway et al., 2004). Typically, eutrophication from nitrogen deposition is caused by anthropogenic sources such as fertilizer and fossil fuel combustion, however, the increased nitrate from ionizing radiation could also cause a eutrophic response.  We addressed specifically whether the increased nitrate rainout from an intense event like a nearby GRB could be a threat to modern terrestrial ecosystems.  In addition, given the likelihood that an event happened in the oxygenated but nitrogen limited past, it could have been an important timely source of biologically available nitrogen in the late-Ordovician.

2. GRB Nitrate Prediction

A nearby GRB represents an extreme case of ionizing radiation making it a good example to use for nitrate producing potential. The Goddard Space Flight Center (GSFC) 2-D atmospheric model was used in previous work to predict the atmospheric effects, including nitrate rainout, of a typical GRB within 2 *kpc* of the earth (Melott et al., 2005, Melott and Thomas, 2009). It considers the atmosphere in two dimensions, latitude and altitude, and varies the angle of incidence of the burst as well as the time of year. It is based on the modern atmosphere, but since the oxygen fraction has generally been 20 ± 10% during the Phanerozoic (Berner et al., 2003, Berner, 2009), it can be used as a semi-quantitative guide to what is expected over this period—with a possible exception during the excursion to very high oxygen levels during the Permo-Carboniferous. For this work we used the highest predicted nitrate rainout rates which were for a September burst striking the earth at 90° N latitude. The burst intensity is one which might be expected to induce a mass extinction, and to occur from a GRB or supernova approximately every 200 Myr (Melott et al., 2005, Piran and Jimenez, 2014, Melott and Thomas, 2011). An important feature of the prediction is that nitrate rainout had a 12 month periodicity and occurred almost exclusively between 30° N and 60° N latitude, making the maximum rainout values correlate well geographically with critical load data compiled for Europe and the US.

The peak month for GRB sourced nitrate rainout rate between 30° N and 60° N latitude occurs 20 months following the burst and is reported in kg of nitrate as $3 \times 10^{-12}$ $kg_{NO3}$ $m^{-2}$ $s^{-1}$(Melott et al., 2005). Since the rainout during winter months is much less ($1 \times 10^{-12}$ $kg_{NO3}$ $m^{-2}$ $s^{-1}$,) we used a conservative estimate for the average monthly value for the year of $2 \times 10^{-12}$ $kg_{NO3}$ $m^{-2}$ $s^{-1}$; which in terms of kg of nitrogen per hectare is $1 \times 10^{-1}$ $kg_N$ $ha^{-1}$ $yr^{-1}$. This localized concentration represents the highest potential value of additional terrestrial nitrogen deposition after an ionizing event. Note that these levels are insignificant for the vast reservoir of the oceans, which in any case have been fixing nitrogen since long before the time we are considering (Raven and Yin, 1998, Navarro-González et al., 2001).

3. Critical Loads

While Thomas and Honeyman already showed that the nitrate rainout is too small to be harmful to amphibians, it is important to establish more generally whether increased nitrate from a GRB would be harmful (Thomas and Honeyman, 2008). While the complexities involved prevent a general predictive model for the evolution of an ecosystem under nitrate addition, the scientific interest in monitoring modern anthropogenic sources of nitrogen deposition, primarily industrial pollution and agricultural inputs, has produced a wealth of data to evaluate its impact. The concept of a critical load is one way that has been developed to systematize the large amount of resulting scientific data. It is defined as "a quantitative estimate of an exposure to one or more pollutants below which significant harmful effects on specified sensitive elements of the environment do not occur according to present knowledge"(Nilsson and Grennfelt, 1988). The critical loads established for total nitrogen deposition include both oxidized and reduced forms, notably nitrate and ammonium which are the primary counter ions in fertilizer. However, the distinction is not typically made in critical load studies since both forms are bioavailable and readily interconverted, therefore, they often give an equivalent eutrophic response. However, their effect on soil pH may differ making it also necessary to evaluate critical loads for acidification.

The use of critical loads to quantify the deposition of pollutants began in the 1980s and has been used extensively in Europe and Canada, and increasingly in the US. The idea is to set a threshold level of input below which there is no known harm on an ecosystem. The metrics used to evaluate harm from eutrophication (excess nutrient deposition) are changes in species composition (which includes decrease in species diversity), plant development, and/or biogeochemistry. In general, the first and most sensitive changes occur with species composition where those adapted for nitrogen starved environments give way to more nitrophilic species, while other changes can include increased mineralization, increased nitrification, nitrate leaching, increased species susceptibility to stress, altered growth patterns, increased N uptake, increased mortality, and increased growth. A synthesis of data for the effects of nitrogen deposition are detailed in Pardo et al., (2011), where the range of compiled critical loads for inland US ecoregions, surface waters and wetlands is between 1–39 *$kg_N$ $ha^{-1}$ $yr^{-1}$*, which are comparable but lower than the critical loads established for Europe (Pardo et al., 2011, Bobbink and Hettelingh, 2011). This establishes the value of 1 *$kg_N$ $ha^{-1}$ $yr^{-1}$* as the lowest critical load for eutrophication reported for those regions.

In addition to eutrophication, acidification is a potential stress that needs further consideration since it can be caused not only by soil mediated processes in response to general excess nutrient nitrogen, but also directly by acid deposition. Unfortunately, the data used for determining critical loads for eutrophication does not distinguish between oxidized or reduced forms of nitrogen and so does not account for direct acidification, specifically by $HNO_3$ (Pardo et al., 2011). The acid neutralizing capacity for systems varies widely and depends on both biotic and abiotic mechanisms. For this paper we will focus on abiotic mechanisms using the critical loads defined by Cinderby et al.(1998) for five soil sensitivity classes which excludes effects from vegetation or climate (Bouwman et al., 2002). By this measure, biological stress is avoided if the buffering capacity of the soil is not exceeded, regardless of species acid tolerance or the mediation of pH due to biotic processes. The critical load, reported in milliequivalents per hectare per year, for the most sensitive soil (class 1) is 25 *meq ha$^{-1}$ yr$^{-1}$* of hydrogen ion , which corresponds to a critical load of 4 *kg$_N$ ha$^{-1}$ yr$^{-1}$* when HNO$_3$ is the source of acid deposition (Bouwman et al., 2002, Cinderby et al., 1998).

In order to establish that the additional nitrate flux from a GRB is not harmful in general to terrestrial ecosystems we will use the lowest defined critical load for nitrogen deposition. Since the critical load for eutrophication is lower than that for acidification, the eutrophication value of 1 *kg$_N$ ha$^{-1}$ yr$^{-1}$* will be used as the maximum nitrogen deposition allowed to be considered not harmful.

4. Implications for ancient ecosystems

It is interesting to briefly address ramifications for ecosystems prior to the dominance of terrestrial biotic nitrogen fixation (BNF), and after the establishment of an oxygenated atmosphere, such as the early Phanerozoic. It has been calculated that a galactic GRB has very likely occurred during the evolution of life on this planet; in fact it is possible that the increased UVB from a GRB could have contributed to the late Ordovician mass extinction (Melott et al., 2004, Melott and Thomas, 2009). It is, however, also possible that increased nitrate rainout during that time could have benefited the invasion of land by plants (Melott et al., 2005). This is due to the fact that in the nitrogen starved world prior to the rise of terrestrial BNF, nitrate rainout could represent a relatively significant source of fixed nitrogen. While marine BNF was underway in the Ordovician, transport to land via the atmosphere would have been insignificant, as most available fixed nitrogen is quickly consumed in the ocean (Schlesinger, 1997). Assuming that lightning input and GRB response in the early Phanerozoic were generally similar to modern values due to largely similar oxygen levels in the atmosphere, we compared the rate for post GRB nitrate deposition to the baseline atmospheric deposition rate. The atmospheric model predicts a localized post-GRB nitrate flux of 1 x 10$^{-1}$ *kg$_N$ ha$^{-1}$ yr$^{-1}$* while the global lightning deposition rate is 5.4 *Tg$_N$ yr$^{-1}$* which corresponds to a localized flux of 1 x 10$^{-1}$ *kg$_N$ ha$^{-1}$ yr$^{-1}$* (Melott et al., 2005, Galloway et al., 2004). We see that the model shows the localized concentrated deposition could have roughly equaled the baseline atmospheric deposition therefore providing up to 50% of the fixed nitrogen in some areas. For comparison, in the modern world even before the boom of anthropogenic nitrogen deposition, terrestrial BNF accounted for 96% of the supply of fixed nitrogen on land at a rate of 120 *Tg$_N$ yr$^{-1}$* while lightning, the primary atmospheric source, accounted for only 4% (Galloway et al., 2004). It should also be mentioned that volcanism and bolide impacts add to atmospheric nitrate deposition, but at a much lower rate of 0.1 *Tg$_N$ yr$^{-1}$*.(Mather et al., 2004, Lewis et al., 1982, Prinn and Fegley, 1987)

Due to the fact that GRB sourced nitrate deposition could double the terrestrial supply of fixed nitrogen for ~5 years and that it is far below critical loads for modern ecosystems, it is reasonable to conclude that any increased supply of fixed nitrogen would be beneficial to nitrogen starved ancient ecosystems. Nitrate would be deposited as the atmosphere recovered to its pre-burst state, allowing plants in ponds and similar environments to receive significant extra nutrient nitrate.

While the possibility of a GRB during the late-Ordovician concerns an oxygenated atmosphere, it has been shown specifically that lightning can fix nitrogen without oxygen in atmospheres dominated by $CO_2$ and $N_2$. (Navarro-González et al., 2001). Therefore, the ionizing radiation of a GRB could be relevant to the Hadean, which suggests a direction for future work.

5. Conclusion

The GSFC atmospheric model predicts nitrate rainout following a GRB at a maximum yearly rate of 0.1 $kg_N\ ha^{-1}\ yr^{-1}$ which is an order of magnitude lower than the most sensitive current critical loads of 1 $kg_N\ ha^{-1}\ yr^{-1}$ for eutrophication and 4 $kg_N\ ha^{-1}\ yr^{-1}$ for acidification. Since the establishment of critical loads is based on present biota, it is reasonable to say that the nitrate rainout from an astrophysical ionizing radiation event would not stress modern pristine ecosystems even when considering the most intense event, namely a GRB, and the most sensitive ecosystems. However, due to modern anthropogenic deposition the estimated amount of global area already exceeding critical loads for acidification is 7%, and for eutrophication is 15% (Bouwman et al., 2002); for these areas GRB nitrate rainout would obviously contribute in a small way to the excess. We conclude that other more catastrophic consequences of a GRB, such as increased UVB radiation, are more important. Due to the low probability of a current nearby GRB (Melott and Thomas, 2011, Piran and Jimenez, 2014), such considerations are more important for understanding the geological past during which it is likely that at least one event has occurred. The UVB effects could have been catastrophic for some species as previously hypothesized for the end-Ordovician extinction (Melott et al., 2004), but if indeed a GRB occurred at that time it is possible the addition of nitrate could have been temporarily beneficial to the transition of plants to land.

6. Acknowledgements
We acknowledge the research support of NASA Exobiology grant NNX14AK22G.